# MRFI Stream Arbitration: An Efficient Arbitration Scheme for NoC with Emerging Interconnect Technology

Rex Lee, Yilei Li

fortelee2016@gmail.com

**Abstract**

An improved version of stream arbitration based on multiband RF interconnect (MRFI) is proposed. Thanks to the simultaneous multiple channel transmitting/receiving feature of MRFI, dynamic bandwidth allocation is achieved in the proposed arbitration algorithm. With dynamic bandwidth allocation, MRFI based arbitration can guarantee 100% channel bandwidth utilization, which is a significant improvement compared with original RF-I based stream arbitration whose channel bandwidth utilization is only around 30%~50%.

1. **Introduction**

   CMOS circuits are the mainstream in current integrated circuits, and its performance keeps evolving with feature size scaling [1-7]. Analog and RF CMOS circuits are usually used for communication [8-10], navigation [11-12], human-machine interface [13-15], and so on; however, RF-based circuits can also be used in processor/memory interface, which is RF-Interconnect (RFI) [16-21]. RFI can provide excellent energy efficiency, and also its RF nature provides potential for flexible arbitration scheme in network-on-chip applications. In [22], stream arbitration based on RF-I was proposed. Compared with other arbitration schemes in emerging interconnect technology, RF-I based stream arbitration has excellent flit transmission latency even with temporal and spatial communication heterogeneity. The authors of [22] also suggest that the efficiency of stream arbitration could be further improved by using dynamic bandwidth allocation. However, with RF-I node which can be tuned to only one frequency channel at one time, it would be very hard to implement dynamic bandwidth allocation.

   On the other hand, the newly proposed multi-band RF interconnect (MRFI) [17-21] node can receive/transmit with multiple frequency channels simultaneously, which enables dynamic bandwidth allocation with minimal overhead. What's more, MRFI has better scalability with technology than original RF-I. Consequently, it is very promising to implement an improved version of stream arbitration with MRFI to achieve superior overall NoC interconnect performance.

   This paper is organized as follows. After introduction in section 1, detailed analysis of RF-I based stream arbitration is described in section 2. The key feature of feature of MRFI that enables dynamic bandwidth allocation is present in section 3. Based on this feature, MRFI stream arbitrationis proposed in section 4. Finally, future works are discussed in section 5.

2. **Limit of RF-I based stream arbitration**

   In RF-I based stream arbitration, each node sends out sub-stream vector and sub-stream vectors form full stream by appending to each other in time domain (trip1). Then each node obtains arbitration result and frequency channel ID from full stream (trip2). Each source-destination pair can only use one frequency channel at a time.

The primary limit of RF-I is that one node can only transmit with one frequency channel at one time. This is due to carrier generation part of RF-I node, which can only generate one frequency for TX/RX at one time.

This implementation limit greatly degrades the resource (channel bandwidth) utilization of RF-I NoC, especially when the traffic pattern is not uniform. One source-destination pair can transfer data with only one frequency channel regardless of whether other channels are idle or not. In many cases other channels are actually idle, and consequently a great portion of channel bandwidth is wasted. This waste is significant when number of active nodes is low. Consider the following case when node0 tries to transmit four flits to node1. Suppose there are four frequency channels for data transfer (each channel can transfer one flit in one cycle), and node0 is the only active node in the network. If node0 could use full bandwidth of the channel (four channels) it could finish data transfer of four flits within one cycle. However, with original RF-I based stream arbitration, node0 can only transfer one flit through one channel in one cycle (while the bandwidth of other three frequency channels are wasted). As a result, it takes four cycles for node0 to finish data transfer. Things become worse when multiple nodes are trying to transfer large amount of data with one node at the same time. Most likely nodes with lower priority have to wait for a long time before they can get channel while the nodes with higher priority are transferring data with low resource utilization (the bandwidth of other idle channels is wasted). Fig. 1 shows the simulated channel utilization in [1]. The average bandwidth utilization of RF-I arbitration is far from efficient with common benchmarks (32% for Blackscholes, 37.8% for EKF-SLAM, 60% for Deblur and 59.5% for Denoise). Unfortunately due to the limit of RF-I, dynamic bandwidth arbitration is not practical for stream arbitration.

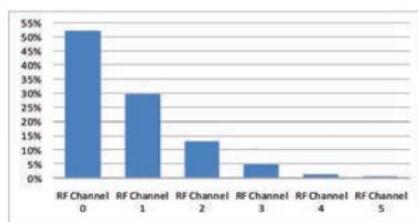
(a) Blackscholes

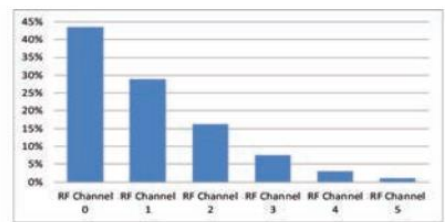
(b) EKF-SLAM

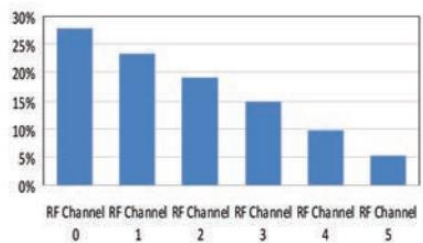
(c) Deblur

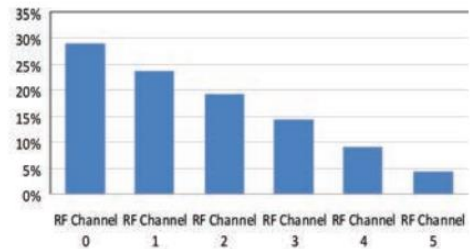
(d) Denoise

Fig. 1 Simulated channel bandwidth utilization of RF-I stream arbitration

Other than the limitation of channel bandwidth utilization, RF-I based stream arbitration may also have issue with its time domain operation of sub-stream appending. Each node needs to finish sub-stream appending within a specific time slot. The length of this time slot is proportional to the travel time of RF signal from one node to another. With process technology scaling, we can expect smaller feature size of transistors and thus smaller distance from one node to another. On the other hand, the speed of RF signal (speed of light) will not scale with technology. This shortens the time slot and it will be hard for nodes in stream arbitration to append sub-stream successfully.

3. **Dynamic bandwidth allocation with MRFI**

The primary factor that prevents stream arbitration from full bandwidth utilization is the fact that RF-I lacks the capacity to work with multiple frequency channels simultaneously. On the other hand, the newly proposed MRFI is able to transmit/receive data with multiple frequency channels at the same time. The architecture of MRFI is shown in Fig. 2.

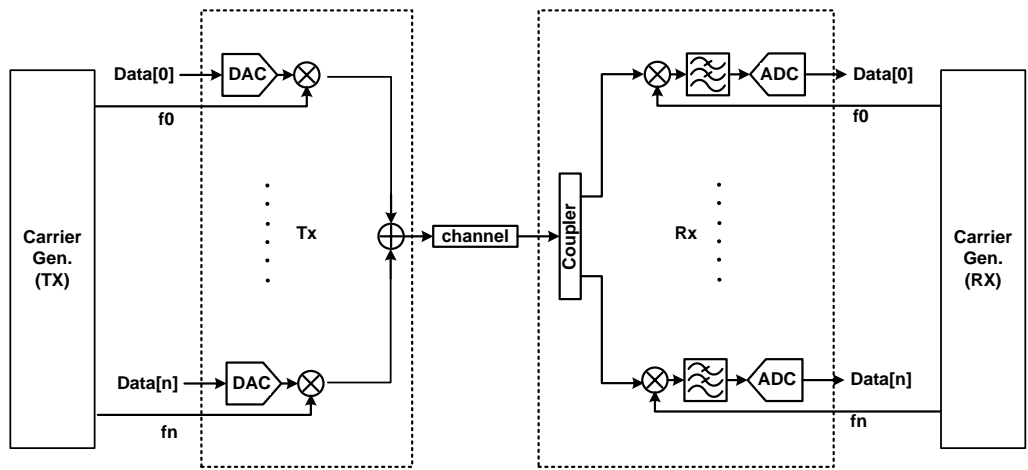

Fig. 2 Architecture of MRFI

The architecture of MRFI is intended to support simultaneous multi-band transmission and receiving. Carrier generation in TX side generates multiple carriers to modulate multiple data streams simultaneously. Those data streams are added together and sent to the channel. In RX side, data streams are demodulated accordingly with multiple carriers. Thus, it is very easy for MRFI to achieve dynamic bandwidth allocation by turning on/off specific modulators. On the other hand, the latency performance of MRFI is similar to that of RF-I, thus low-latency arbitration is still available for MRFI-based arbitration.

An exemplary circuit-level simulation result is shown in Fig. 3. In this example, two nodes with three channels are sharing one channel. In time period t1, only node1 is active, and thus it uses all channel bandwidth to communicate with node2. In time period t2, both node1 and node2 are active. Node1 uses 66% of channel bandwidth and node2 uses 33% of channel bandwidth to communicate with each other. In time period t3, only node2 is active, and node2 uses all bandwidth of the channel to transfer data to node1. Clearly, dynamic allocation of bandwidth keeps channel bandwidth utilization at 100% at all time.

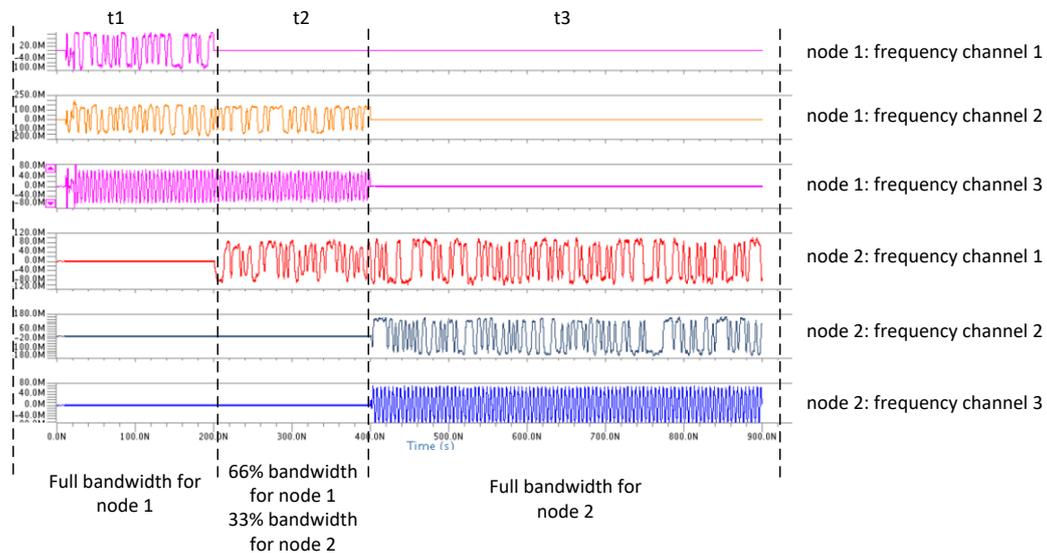

Fig. 3 Simulation waveform of dynamic bandwidth allocation with MRFI

## 4. MRFI-Based Stream Arbitration

MRFI can implement stream arbitration with dynamic bandwidth allocation. In addition, the sub-stream appending can be accomplished in frequency domain rather than in time domain and therefore MRFI-based stream arbitration has better scalability than original RFI-based stream arbitration.

Every node in MRFI-based stream arbitration has specific arbitration frequency channel. Each node sends out sub-stream vector in its arbitration frequency channel. The sub-stream vector format is similar to original stream arbitration.

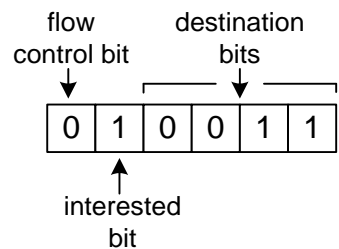

Fig. 4 Substream vector in MRFI-based arbitration

When a node is available for receiving data, it sets flow control bit to '0'. When a node wants to use channel to transfer data to another node (destination node) it sets interested bit to '1' and destination bits to the ID of the destination node (with $n$ nodes in NoC, the total destination bits will be $\log_2 n$).

In arbitration, every node sends its sub-stream vector into specific arbitration frequency channel. After a very short latency (which equals interconnect distance divided by speed of light) every node is able to receive sub-stream vectors in all arbitration channels, and thus full stream vector is obtained by every node (Fig. 5). Unlike time-domain appending of sub-stream vectors in original stream arbitration, the MRFI stream arbitration uses frequency-domain appending of sub-stream vectors. The bandwidth required for MRFI stream arbitration is well below the capacity of mainstream technology. For example, with 2-GHz cycle frequency and 16 nodes in NoC, each node sends out 6 bits in one cycle in one channel. With QAM-64 modulation every arbitration channel needs 2 GHz bandwidth to send out 6 bits in one cycle. According to simulation, 4 GHz channel spacing is sufficient for a low bit-error-rate of $10^{-12}$. Consequently, it takes 64 GHz bandwidth to support arbitration of 16 nodes. Hence mainstream technology (e.g., 32 nm technology with cut-off frequency $f_t$ of 350 GHz [1]) can handle arbitration bandwidth with much margin. Furthermore, with higher $f_t$ in advanced technology, more frequency channels are available for arbitration to support more nodes, so feature size scaling in fact benefits MRFI stream arbitration. This is in contrast to the original stream arbitration where feature size scaling may induce too short time slot for sub-stream vector appending.

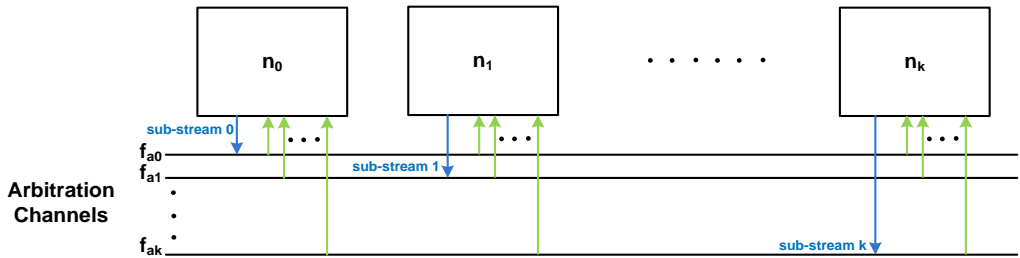

Fig. 5 Arbitration with *k* nodes: each node transmits sub-stream in specific frequency channel and receives sub-streams from all channels

After each node receives full stream (within one cycle), each node extracts arbitration result from the stream. For nodes interested in transmitting data, they decide which channels (probably more than one channel) are allocated to them for data transfer from stream. For nodes that are available for receiving data, they obtain the channel IDs that are used to receive data from source nodes. Destination nodes will only talk with one source node with the highest priority among all source nodes that are sending request to it. However it is possible to extend the arbitration algorithm so that one destination node can talk with multiple source nodes later.

### 4.1 MRFI-based stream arbitration algorithm

Suppose there are *k* nodes for arbitration (and thus *k* frequency channels for arbitration, $f_a[1]$~$f_a[k]$) and *m* data channels ($f_d[1]$~$f_d[m]$) are available for data transfer. Node with higher priority uses arbitration frequency channel with lower ID index (i.e., $f_a[1]$ is used by node with highest frequency and $f_a[k]$ is used by node with lowest priority). For any node $n_i$, its sends out sub-stream vector to its specific arbitration frequency channel, and receives full stream from all arbitration channels. Let's assume this node $n_i$ is interested in transferring data and also available for receiving data.

For TX side of $n_i$, it first needs to find out whether the destination node is available for receiving data by checking the flow control bit of destination node. If the flow control bit of destination node is '0' (means it is available for receiving data), then $n_i$ needs to find out what channels are allocated to it. Suppose there are $q$ source-destination pairs to share the channel bandwidth and there are $p$ source-destination pairs that have higher priority than it (and no source node with higher priority is competing for the same destination with $n_i$). Then any data channel with ID of $p+1+jq$, where $j$ is any natural number that satisfies $p+1+jq<m+1$, is allocated to node $n_i$ for transmitting data (Fig. 6). By allocating channel bandwidth in this manner, the channel bandwidth utilization is guaranteed to be 100%. When there are fewer active nodes, i.e., $q$ is small, then each active node is allocated with more channels (i.e., more bandwidth) as more $j$ can satisfy $p+1+jq<m+1$. When number of active nodes is large, then nodes with high priority evenly share the channel bandwidth.

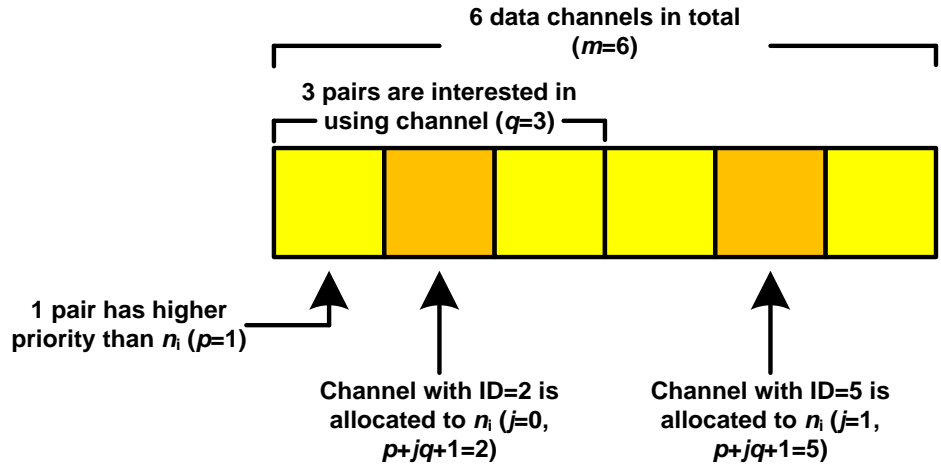

Fig. 6 Dynamic channel bandwidth allocation with 6 data channels ($m$=6) and 3 active nodes ($q$=3). Algorithm allocates two channels to each active node.

In order to find out number of pairs with higher priority ($p$) and total number of pairs that are interested in using channel ($q$), node $n_i$ parses sub-stream vectors start from frequency channel with highest priority $f_a[1]$. For TX, there exists $p\_t$ as the number of pairs with higher priority than node $n_i$ for TX. For RX, there also exists $p\_r$ as the number of pairs with higher priority than the node that tries to access node $n_i$. The algorithm is close to original stream arbitration algorithm when finding $p\_t$, $p\_r$ and $q$ by manipulating *flowControl* bit of status vector. Define an array *TX_CH*[$i$] where $i$=[1…$m$] for channels to use for data transfer. *TX_CH* is reset to all '0' before arbitration starts. If *TX_CH*[$i$] is '1' then node $n_i$ will use this channel for data transfer. After $p\_t$ and $q$ are obtained, set all *TX_CH*[$p\_t+1+jq$]='1', where $j$ is any natural number that $p\_t+1+jq<m+1$. Similarly, define an array *RX_CH*[$i$] where $i$=[1…$m$] for channels to use for data transfer. *RX_CH* is reset to all '0' before arbitration starts. If *RX_CH*[$i$] is '1' then node $n_i$ will use this channel for data receiving. $RX\_CH[p\_r+1+jq]$ is set to '1', where $j$ is any natural number that $p\_r+1+jq<m+1$. The whole algorithm is quite simple and straightforward, and it can be finished within one or two cycles depending on technology. On the other hand, the MRFI stream arbitrationonly takes one trip to finish. Thus the arbitration time is similar to original stream arbitration.

**Input**: Stream: *flowControl*[1…K], *interested*[1…K], *destination*[1…K], where *K* is the number of RF nodes; the total number of channels *M*; priority table *priorityMap*[1…K] which maps priority to corresponding node's ID; this node's ID *node_id* and its priority *priority*.

**Output:** TX channel status table *TX_CH*[1…M] and RX channel status table *RX_CH*[1…M]

for (i=1;i<K+1;i++) {

if (*interested*[*priorityMap*[i]]== 1 and *flowControl*[*destination*[*priorityMap*[i]]]== 0){

*flowControl*[*destination*[*priorityMap*[i]]]=1;

if (i==*priority*)

*p_t=q*;     //node with higher priority

if (*destination*[*priorityMap*[i]]==*node_ID*)

*p_r=q*;

*q*++;

if (*q == M*)

break;

        }

}

for (i=1;p_t+i*q<M+1;i++) {

TX_CH[p_t+i*q]=1;

}

for (i=1;p_r+i*q<M+1;i++) {

RX_CH[p_r+i*q]=1;

}

### 4.2 Priority adjustment

The priority of each node is easily adjusted by changing the arbitration frequency channel that it uses (and adjusting the *priorityMap* table accordingly). Arbitrary priority adjustment scheme is supported by MRFI-based arbitration. Simple rotary priority adjustment has good result in terms of fairness, and more complicated scheme is also possible. Note that since the stream information is open to every node in arbitration (i.e., every node in arbitration has perfect information about the game), so priority adjustment based on optimal strategy may help to achieve global fairness.

**4.3 An example of MRFI-based stream arbitration**

We assume MRFI-based stream arbitration takes only one cycle to finish. In this example, we have four nodes $n_1$~$n_4$ with four arbitration frequency channels $f_a[1]$~$f_a[4]$ and four data channels $f_d[1]$~$f_d[4]$. Every data channel can transfer one flit in one clock cycle. Assume simple scenario of static priority and $n_1$ has highest priority while node $n_4$ has the lowest priority. Node $n1$ has four flits to transmit to $n_2$, and node $n_3$ has two flits to transfer to $n_2$. Node $n_1$ and $n_2$ make request to transfer in cycle 0. In cycle 1, node $n_4$ makes request to transfer two flits to $n_1$. The RX buffers of all nodes are always available to receive data.

In cycle 0, $n_1$ sends out sub-stream vector 0101, $n_2$ sends out sub-stream vector 0000, $n_3$ sends out sub-stream vector 0101 and $n_4$ sends out sub-stream vector 0000 (Fig. 7). Node $n_1$ executes algorithm upon full stream and finds out that total one source-destination pair will use channel ($q$=1) and zero source-destination pair has higher priority than it ($p$=0). Thus, $p+1+jq$ can be 1 ($j$=0), 2 ($j$=1), 3 ($j$=3) and 4 ($j$=4). Consequently node $n_1$ will use all four channels to transmit data to node $n_2$. RX of node $n_1$ does not detect any node that sends request to talk with it so all channels in RX side of node $n_1$ are turned off. RX of node $n_2$ uses algorithm to find out that $n_1$ will use four channels to talk with it and thus turn on four RX channels accordingly in next data transfer cycle. Node $n_3$ finds out that node $n_1$ with higher priority is also competing for destination node $n_2$, thus it loses arbitration and will turn off all TX channels. RX side of node $n_3$ and TX/RX side of node $n_4$ stay idle as there is no request.

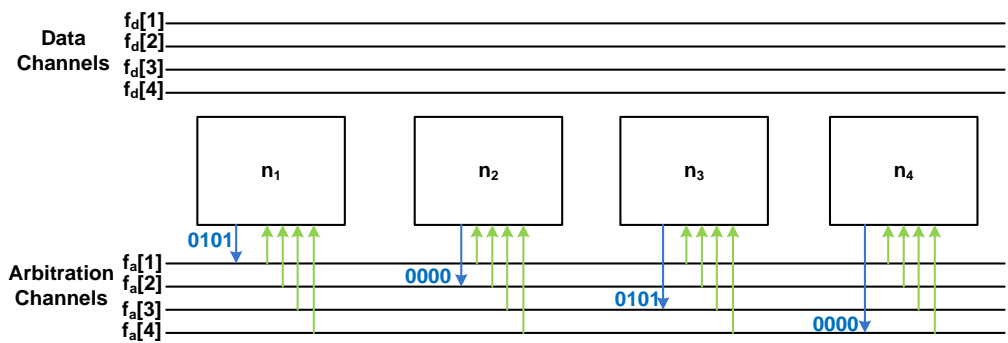

Fig. 7 Cycle 0 of arbitration

In cycle 1, $n_1$ uses all four data channels to send data to $n_2$ and finishes data transfer of four flits within one cycle. Node $n_3$ tries the second time to reach node $n_2$. Node $n_4$ sends request to send data to $n_1$. Node $n_1$ sends out sub-stream vector 0000, $n_2$ sends out sub-stream vector 0000, $n_3$ sends out sub-stream vector 0101 and $n_4$ sends out sub-stream vector 0100 (Fig. 8). Node $n_3$ executes algorithm upon full stream and finds out that total two source-destination pairs will use channel ($q$=2) and zero source-destination pair has higher priority than it ($p$=0). Thus, $p+1+jq$ can be 1 ($j$=0) and 3 ($j$=1). Consequently node $n_3$ will use data channel 1 and data channel 3 to transmit data to node $n_2$. RX of node $n_2$ uses algorithm to find out that $n_3$ will use data channel 1 and data channel 3 to talk with it and thus turn on RX channels accordingly in next data transfer cycle. Node $n_4$ finds out that total two source-destination pairs will use channel ($q=2$) and one source-destination pair has higher priority than it ($p=1$). Thus, $p+1+jq$ can be 2 ($j=0$) and 4 ($j=1$). Consequently node $n_4$ will use data channel 2 and data channel 4 to transmit data to node $n_1$. RX of node $n_1$ detects that node $n_4$ will transmit data to it through data channel 2 and data channel 4, and tunes to data channel 2 and data channel 4 accordingly in next data transfer cycle.

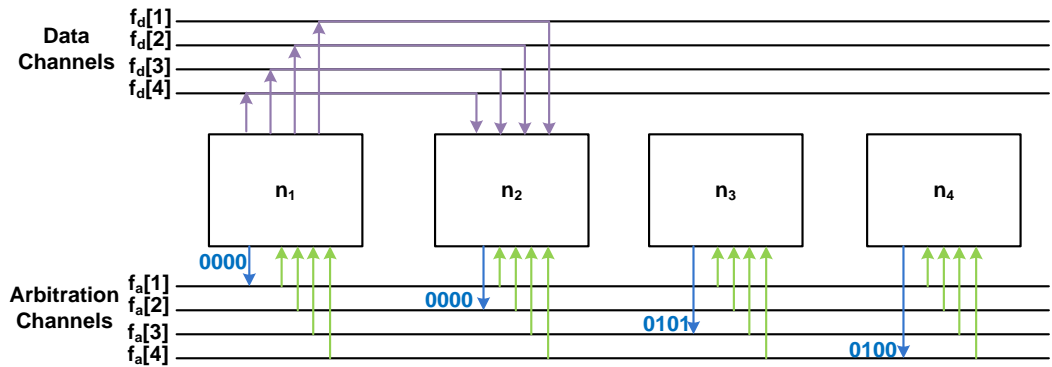

Fig. 8 Cycle 1 of arbitration

In cycle 2, node $n_3$ uses data channel 1 and data channel 3 to transfer two flits to node $n_2$, and node $n_4$ uses data channel 2 and data channel 4 to transfer two flits to node $n_1$. The data transfer finishes within one cycle (Fig. 9).

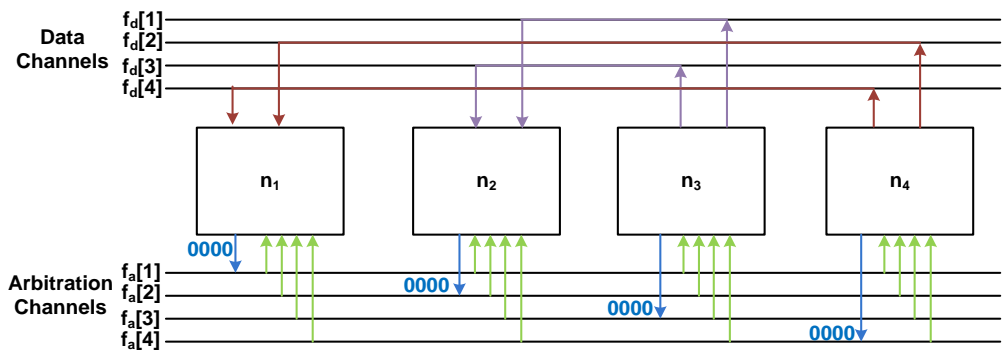

Fig. 9 Cycle 2 of arbitration. Data transfer finishes.

Clearly, in the example, MRFI-based arbitration uses two cycles to finish total data transfer of eight flits. The channel bandwidth utilization is 100% and longest wait time is one cycle (node $n_2$). On the other hand, if RF-I based original stream arbitration, where one source-destination pair can only use one channel at one time, is used in this case, then it takes much longer time to finish data transfer. Node $n_1$ will take four cycles to finish transmitting four flits, and node $n_3$ needs to wait four cycles before it can start. Then node $n_3$ will take two cycles to finish transferring two flits. It takes six cycles to finish transferring of eight flits and average channel bandwidth utilization is only 33%. The longest wait time is four cycles (node $n_3$). The comparison of results of MRFI based stream arbitration and RF-I based stream arbitration is shown in Table 1 (in comparison we also assume RF-I based stream arbitration only takes one cycle to finish).

Table 1. Comparison of MRFI based stream arbitration and RF-I based stream arbitration

|  | MRFI based | RF-I based |
|---|---|---|
| Total flits for transfer | 8 | 8 |
| Total cycles for transfer | 2 | 6 |
| Longest wait cycle | 1 | 4 |
| Bandwidth utilization | 100% | 33% |

Clearly, MRFI based stream arbitration achieves significant improvement in total latency (3 cycles vs. 8 cycles) and bandwidth utilization (100% vs. 33%) over original RF-I based stream arbitration.

**5. Summary and Future Works**

An improved version of stream arbitration based on MRFI is proposed. Thanks to the simultaneous multiple channel transmitting/receiving feature of MRFI, dynamic bandwidth allocation is achieved in the proposed arbitration algorithm. With dynamic bandwidth allocation, MRFI based arbitration can guarantee 100% channel bandwidth utilization, which is a significant improvement compared with original RF-I based stream arbitration whose channel bandwidth utilization is only around 30%~50%. With higher bandwidth utilization, shorter latency and wait time of flit transfer are achieved (2~3X improvement). Furthermore, unlike RF-I based stream arbitration which appends sub-stream vectors in time domain, MRFI based stream arbitration appends sub-stream vectors in frequency domain. As a result, the proposed arbitration has better scalability with technology feature size than original stream arbitration.

The MRFI based stream arbitration, however, still has more potential to exploit. First, in the proposed scheme, one destination node only communicates with one source node. In fact, MRFI supports one destination node to communicate with multiple source nodes. It will be interesting to find out how multiple-point receiving will benefit the whole system. Second, broadcast of source node is also supported by MRFI. It may be useful for certain applications. Finally, an optimal priority adjustment scheme can be investigated to guarantee global fairness, probably based on game theory.